\documentclass[english,two column]{revtex4-1}
\usepackage[LGR,T1]{fontenc}
\usepackage[latin9]{inputenc}
\usepackage{amsmath}
\usepackage{amssymb}
\usepackage{graphicx}
\usepackage{hyperref}
\hypersetup{breaklinks,
colorlinks=true,%
citecolor=blue,%
filecolor=black,%
linkcolor=blue,%
urlcolor=blue}

\makeatletter

\ProvideTextCommand{\~}{LGR}[1]{\char126#1}

\makeatother

\usepackage{babel}
\begin{document}
\title{Dynamical structure factor and a new method to measure the pairing gap in two-dimensional attractive Fermi-Hubbard model}

\author{Huaisong Zhao$^{1}$}

\author{Feng Yuan$^{1}$}
\email{yuan@qdu.edu.cn}
\author{Tianxing Ma$^{2}$}
\email{txma@bnu.edu.cn}
\author{Peng Zou$^{1}$}
\email{phy.zoupeng@gmail.com}
\affiliation{$^{1}$Centre for Theoretical and Computational Physics, College of Physics, Qingdao University, Qingdao 266071, China}
\affiliation{$^{2}$School of Physics and Astronomy, Beijing Normal University, Beijing 100875, China}
\begin{abstract}
The measurement of the pairing gap plays an essential role in studying the physical properties of superconductors or superfluids. We develop a strategy for measure the pairing gap through the dynamical excitations. With the random phase approximation (RPA), the dynamical excitations of a two-dimensional attractive Fermi-Hubbard model are studied by calculating the dynamical structure factor. Two distinct collective modes are investigated: a Goldstone phonon mode at the transferred momentum ${\bf q}=\left[0,0\right]$ and a roton mode at ${\bf q}=\left[\pi,\pi\right]$. The roton mode demonstrates a sharp molecular peak in the low-energy regime. Remarkably, the area under the roton molecular peak scales with the square of the pairing gap, which persists even in three-dimensional and spin-orbit coupled (SOC) optical lattices. This result provides a potential strategy to measure the pairing gap of lattice systems experimentally by measuring the dynamical structure factor at ${\bf q}=\left[\pi,\pi\right]$.

\end{abstract}
\maketitle

\section{Introduction}
The superfluid state of quantum many-body Fermi gases has a nonzero pairing gap (order parameter) originating from the Cooper pairing. Establishing a convenient method to measure the pairing gap is essential to understanding many-body pairing phenomena and dynamical excitations. Currently, the pairing gap is mainly obtained by virtue of various excitation processes, such as the momentum-resolved photo-emission spectroscopy \cite{Feld11,Stewart2008} and the radio-frequency spectroscopy \cite{Frohlich2011,Chin2004,Sommer2012}. However, these methods face great difficulties in systems with complex band structure \cite{Zhai2015,Cheuk2012,Wang2012,Wang2021,Wu2013,Han2023}. Notably,
 the dynamical excitations provide an important method for studying the pairing correlations. These excitations are emerged through the dynamical structure factor which can be experimentally measured through the two-photon Bragg spectroscopy  \cite{Veeravalli08,Hoinka17,Dyke2023,Biss2022,Senaratne2022,Li2022,Pagano2014}.

 In continuous Fermi gases, the collective modes are typically probed at a small transferred momentum ${\bf q}$, whereas the single-particle excitations of the unpaired and paired atoms emerge at a relatively larger ${\bf q}$. Specifically, the excitations of the paired atoms correspond to the bosonic molecular excitations.
In 2008, Vale {\it et al.} studied the single-particle excitations at a large transfer momentum $q{\gg}k_{F}$. They found that the molecular scattering peak gains increasing spectral weight when tuning the interaction from the Bardeen-Cooper-Schrieffer (BCS) regime to Bose-Einstein-Condensate (BEC) regime, being different from the behaviour of atomic scattering \cite{Veeravalli08}. Later, they investigated the Goldstone phonon mode and pair-breaking excitation at a small transferred momentum, revealing interaction-induced suppression of the sound speed \cite{Hoinka17}. Then the bosonic molecular excitations is found to be
sensitive to variations in the pairing gap \cite{Dyke2023}. Biss {\it et al.} experimentally studied the phonon dispersion across the BCS-BEC crossover using the Bragg spectroscopy \cite{Biss2022}.
 The dynamical structure factors of three-dimensional (3D) Fermi gases have been extensively studied theoretically \cite{Combescot06,Combescot2006,Zou10,Zou16,Zou18,Hu18,Zou2021,Kuhnle10,Watabe10}. For two-dimensional (2D) superfluid Fermi gases, the recent experiments using the two-photon Bragg spectroscopy measured the dynamical structure factor across the entire interaction strength \cite{Sobirey2022}. Notably, the exact quantum Monte Carlo (QMC) method was employed by Vitali {\it et al.} to investigate the spectral weight redistribution between the molecular and atomic excitations \cite{Vitali17}. For other low-dimensional Fermi gases, several theoretical works had investigated the dynamical excitations with dynamical structure factor \cite{Zhao2020,Gao2023}.

As to the discrete ultracold atomic gases, the optical lattice systems generated by superimposing orthogonal standing waves provide a great platform for simulating the physics in crystalline systems \cite{Bloch2008,Wu2016}. These systems are described by the Bose-Hubbard or Fermi-Hubbard models \cite{Greiner2002,Spielman2008,Thomas2017,Jrdens08,Schneider08,Greif13,Hart15,Parsons16,Cheuk16,Koepsell2021,Boll16,Brown17,Arovas2022}. Several theoretical groups have studied the attractive Fermi-Hubbard model due to its relevance to the strongly correlated systems in condensed matter physics \cite{Scalettar89,Kyung01,Honerkamp2004,Mondaini2015,Cocchi16,Strohmaier07,Ho04,Moreo07,Gukelberger16,Paiva04,Shenoy2008}. Experimentally, this model in cold atoms has been realized \cite{Mitra18,Hackermuller10,Peter20,Gall2020,Schneider12,Hartke2023}. To date, no experimental measurement of the dynamical structure factor has been reported on a 2D optical lattice.
Based on the QMC simulations for the half-filled attractive Fermi-Hubbard model on a square optical lattice, Vitali {\it et al.} numerically investigated the dynamical structure factor along the high-symmetry Brillouin zone (BZ) directions \cite{Vitali2020}, revealing both the low-energy two collective modes and higher-energy single-particle excitations. However, the relationship between the roton mode and the pairing gap on an optical lattice remains unclear.

In this paper, we theoretically investigate the doping dependent dynamical excitations of 2D Fermi superfluid on an optical lattice from the weak to the intermediate coupling regime. Through detailed analysis of the dynamical structure factor, the main characteristics of both the collective modes and the single-particle excitations are demonstrated,
particularly focusing on the roton molecular peak at the momentum ${\bf q}=\left[\pi,\pi\right]$. This roton molecular peak provides a strategy for measuring the pairing gap in doped systems, namely, the square of the pairing gap is proportional to the area under this peak. Notably, this strategy can be generalized to the SOC Fermi gases on an optical lattice, where the complex band structure makes it difficult to directly measure the pairing gap \cite{Zhao2023}.

This paper is organized as follows. In Sec. \ref{modelH}, we use the equations of motion of the Green's functions to solve the 2D Fermi-Hubbard model in mean field approximation, and
self-consistently obtain the chemical potential and pairing gap. In Sec. \ref{DSF}, we introduce how to calculate the dynamical structure factor with random phase approximation (RPA).  We display results of dynamical structure factor at half-filling and compare with the QMC results in Sec. \ref{halfres}. In Sec. \ref{awhalf}, we present results when the system is away from half-filling, and discuss the hopping dependence of the sound speed and the dynamical excitations at a transferred momentum ${\bf q}=\left[\pi,\pi\right]$. We check the doping dependence of dynamical structure factor in Sec. \ref{Dopingdsf}. Finally we give our conclusions and acknowledgment, and provide some calculation details in the appendix.

\section{Model and Hamiltonian}
\label{modelH}
An attractive Fermi-Hubbard model in 2D square optical lattices can be described by a Hamiltonian in spatial representation as follows:
\begin{eqnarray}\label{Humodel2}
H =&-&t\sum_{<ij>}C_{i\sigma}^{\dagger}C_{j\sigma}-\mu\sum_{i}C_{i\sigma}^{\dagger}C_{i\sigma}\notag\\
&-&U\sum_{i}C_{i\uparrow}^{\dagger}C_{i\downarrow}^{\dagger}C_{i\downarrow}C_{i\uparrow},
\end{eqnarray}
where $\left<ij\right>$ means the nearest-neighbor sites of lattice. $C_{i\sigma}^{\dagger}(C_{i\sigma})$ is the creation (annihilation) operator of a particle with spin $\sigma$, hopping energy $t$ and chemical potential $\mu$ at site $i$. The Hubbard energy $U>0$ is the strength of on-site two-body attraction interaction. In the following discussions, $U$ is set to be the unit energy,  while the lattice length $a_{0}$ is used as unit length. Within the mean field theory, the four-operators interaction Hamiltonian can be dealt into a two-operators one with the definition of pairing gap  $\Delta=U\left<C_{i\downarrow}C_{i\uparrow}\right>$.  The pairing gap $\Delta$ can be chosen to be a real number in the ground state. Then the above Hamiltonian is displayed into a mean field one, whose expression in momentum space reads
\begin{eqnarray}\label{tjbm}
H_{\rm MF}&=&\sum_{{\bf k},\sigma}\xi_{\bf k}C^{\dagger}_{{\bf k}\sigma}C_{{\bf k}\sigma}\notag\\
&-&\sum_{{\bf k}}(\Delta^{*}C_{{\bf k}\downarrow}C_{-{\bf k}\uparrow}+H.c.)+\frac{|\Delta|^{2}}{U},
\end{eqnarray}
where $\xi_{\bf k}=-Zt\gamma_{\bf k}-\mu$ and $\gamma_{\bf k}=\left(\cos{k_{x}}+\cos{k_{y}}\right)/2$. The nearest lattice number satisfies $Z=4$ for 2D square lattice.

The above mean field Hamiltonian can be solved by the equations of motion of the Green's functions. Here we define the diagonal Green's function $G({\bf k},\tau-\tau{'})=-\left\langle T_{\tau} C_{{\bf k}\sigma}(\tau)C^{\dagger}_{{\bf k}\sigma}(\tau{'})\right\rangle$  and off-diagonal one $\Gamma^{\dagger}({\bf k},\tau-\tau{'})=-\left\langle T_{\tau} C^{\dagger}_{-{\bf k}\uparrow}(\tau)C^{\dagger}_{{\bf k}\downarrow}(\tau{'})\right\rangle$, respectively. The diagonal Green's function is related to the normal particle density and the off-diagonal Green's function is related to the singlet Cooper pairing information. Their expressions in momentum and energy representation are given by
\begin{subequations}\label{bcsgreen}
 \begin{eqnarray}
G(\bf{k},\omega)&=&\frac{1}{2}\left(\frac{1+{\xi}_{\bf{k}}/E_{\bf{k}}}{\omega-E_{\bf{k}}}+
\frac{1-{\xi}_{\bf{k}}/E_{\bf{k}}}{\omega+E_{\bf{k}}}\right)\\
\Gamma^{\dagger}\left(\bf{k},\omega\right)&=&\frac{\Delta^{*}}{2E_{\bf{k}}}
\left(\frac{1}{\omega-E_{\bf{k}}}-\frac{1}{\omega+E_{\bf{k}}}\right),
\end{eqnarray}
\end{subequations}
where $E_{\bf{k}}=\sqrt{{\xi}^2_{\bf{k}}+{|\Delta|^{2}}}$ is the quasiparticle spectrum.
The chemical potential $\mu$ and pairing gap $\Delta$ are determined by self-consistently solving the density equation and pairing gap equation
\begin{eqnarray}\label{twoequtions}
n &=&\frac{1}{2}\sum_{{\bf k}}\left(1-\frac{{\xi}_{\bf{k}}}{E_{\bf{k}}}\right){\rm tanh}\left(\frac{ E_{\bf{k}}}{2T}\right),\nonumber\\
1&=& \frac{U}{N}\sum_{{\bf k}} \frac{1}{2E_{{\bf k}}}{\rm tanh}\left(\frac{E_{{\bf k}}}{2T}\right),
\end{eqnarray}
We have set Boltzmann constant $k_B=1$, and will consider a typical low temperature $T/U=0.01$ (close to zero) in the following.

\begin{figure}[h!]
\includegraphics[scale=0.5]{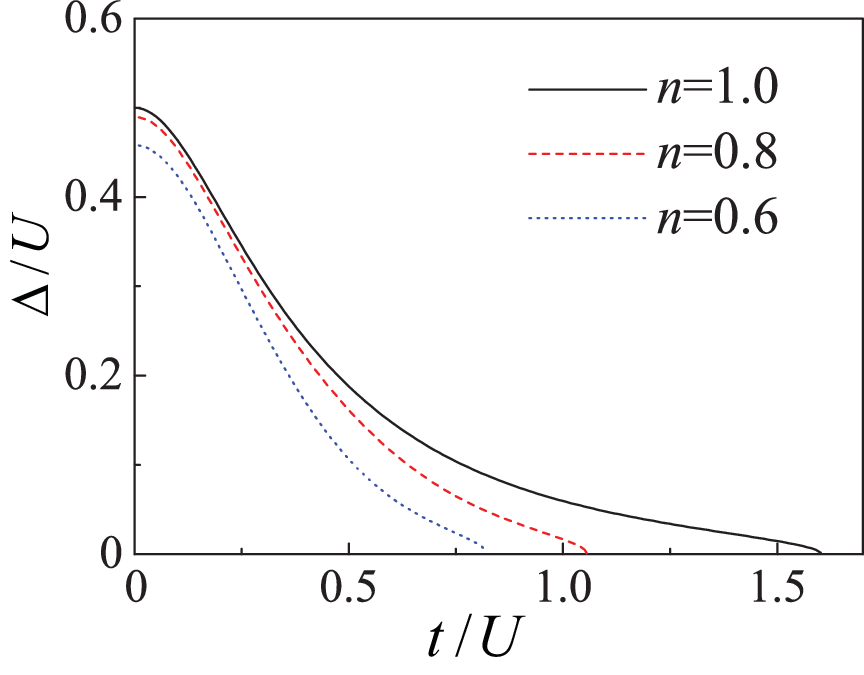}
\caption{Pairing gap $\Delta$ as a function of $t$ for $n = 1.0$ (black solid line), $n = 0.8$ (red dashed line) and $n = 0.6$ (blue dotted line).
\label{fig1}}
\end{figure}

 Generally, the pairing gap $\Delta$ exhibits an inverse dependence on the hopping strength $t$. We plot $\Delta$ as a function of $t$ for different particle densities $n$ in Fig. \ref{fig1}. When $t$ is large, $\Delta$ asymptotically approaches zero, indicating a phase transition from a superfluid to a normal state. Furthermore, $\Delta$ is significantly suppressed as $n$ decreases. It should be noted that $\Delta$ under the mean-field theory is overestimated compared with the corresponding QMC results \cite{Vitali2020}. For example, at half-filling ($n=1.0$) with $t/U=0.25$, the mean-field theory yields $\Delta/U=0.345$ while the QMC predicts $\Delta/U=0.1825$. Experimentally, the formation and spatial ordering of non-local fermion pairs in an attractive Fermi-Hubbard system are directly observed using two-species degenerate $^{40}$K atomic gases \cite{Hartke2023}.

\section{Dynamical structure factor and random phase approximation}
\label{DSF}
In this section, we introduce the main idea of the random phase approximation (RPA). The RPA provides an approach to systematically investigate the dynamical excitations by incorporating quantum fluctuations beyond the mean-field theory.

In a superfluid system, there are four different densities, with the normal spin-up/spin-down $\hat{n}_{1}=\left <C^{\dagger}_{i\uparrow}C_{i\uparrow}\right>$, $\hat{n}_{2}=\left<C^{\dagger}_{i\downarrow}C_{i\downarrow}\right>$, and the two anomalous densities describing Cooper pairing, $\hat{n}_{3}=\left<C_{i\downarrow}C_{i\uparrow}\right>$, $\hat{n}_{4}=\left<C^{\dagger}_{i\uparrow}C^{\dagger}_{i\downarrow}\right>$. These four densities are coupled with each other through the finite interaction, where any perturbation in one density induces the correlated fluctuations in other densities. Within the linear response theory, the small external perturbation potential $V_{\rm ext}$ and density fluctuations $\delta n$ are connected with each other through the full response function $\chi$, namely, $\delta n=\chi V_{\rm ext}$.

 The mean-field theory neglects the contribution from the fluctuation term of interaction Hamiltonian, failing to predict the dynamical excitations in interacting systems. In order to incorporate the quantum fluctuations \cite{Liu2004,He2016,Ganesh2009}, the RPA has been proven to be a reliable method for calculating $\chi$ beyond the mean-field level. In a 3D BCS-BEC crossover Fermi superfluid, the dynamical excitations obtained from RPA theory even quantitatively agree well with the experimental results \citep{Biss2022,Zou10,Zou18}.
 Similarly, the 2D RPA incorporating the density fluctuations can qualitatively reproduce the corresponding QMC data \cite{Zhao2020}. These results indicate that the RPA approach can provide qualitatively reliable predictions for the dynamical excitations of a 2D lattice Fermi superfluid.

The main idea of the RPA is to treat fluctuation Hamiltonian as part of an effective external potential. This theory establishes the relationship between the full response function $\chi$ beyond mean-field level and its mean-field response function $\chi^0$,
 \begin{eqnarray}\label{chi}
 \chi({\bf q},i\omega_{n})=\frac{\chi^{0}({\bf q},i\omega_{n})}{\hat{1}+\chi^{0}({\bf q},i\omega_{n})UG}.
\end{eqnarray}
Here $G=\sigma_{0}\otimes\sigma_{x}$ is a direct product of unit matrix $\sigma_{0}$ and Pauli matrix $\sigma_{x}$.

The numerical calculation of mean-field response function $\chi^0$ is easy to gain, and its expression is given as
\begin{eqnarray}\label{matrix}
\chi^{0}({\bf q},i\omega_{n})=\left[
\begin{array}{cccccc}
&\chi^{0}_{11}&\chi^{0}_{12}&\chi^{0}_{13}&\chi^{0}_{14}\\
&\chi^{0}_{21} &\chi^{0}_{22}&\chi^{0}_{23}&\chi^{0}_{24}\\
&\chi^{0}_{31}&\chi^{0}_{32}&\chi^{0}_{33}&\chi^{0}_{34}\\
&\chi^{0}_{41}&\chi^{0}_{42}&\chi^{0}_{43}&\chi^{0}_{44}\\
\end{array}
\right].
 \end{eqnarray}
 The dimension of $\chi^0$ demonstrates the coupling situation among the four density channels.
 These 16 matrix elements are obtained through the corresponding density-density correlation functions derived from the previously defined Green's functions. Due to all possible symmetries of system, only 6 of these matrix elements are independent, i.e., $\chi^{0}_{11}=\chi^{0}_{22}$,
 $\chi^{0}_{12}=\chi^{0}_{21}=-\chi^{0}_{33}=-\chi^{0}_{44}$,
 $\chi^{0}_{31}=\chi^{0}_{32}=\chi^{0}_{14}=\chi^{0}_{24}$,
 $\chi^{0}_{41}=\chi^{0}_{42}=\chi^{0}_{13}=\chi^{0}_{23}$.
 The symmetry of the matrix is closely related to the symmetry of the Green's functions, such as $\Gamma^{\dagger}\left(\bf{k},\omega\right)=\Gamma\left(\bf{k},\omega\right)=\Gamma\left(\bf{k},-\omega\right)$, which leads to $\chi^{0}_{12}=\chi^{0}_{21}$, where $\chi^{0}_{12}=-\left\langle T_{\tau} C^{\dagger}_{\uparrow}({\bf r},\tau)C_{\uparrow}({\bf r},\tau)C^{\dagger}_{\downarrow}({\bf r'},\tau{'})C_{\downarrow}({\bf r'},\tau{'})\right\rangle$. Based on Wick's theorem, $\chi^{0}_{12}=-\Gamma^{\dagger}({\bf r}-{\bf r'},\tau-\tau')\Gamma({\bf r'}-{\bf r},\tau'-\tau)$. Their expressions are listed in the final appendix of this paper.

The total density response function $\chi_n$ is defined by $\chi_n \equiv\chi_{11}+\chi_{12}+\chi_{21}+\chi_{22}$, its expression is given as
\begin{eqnarray}\label{ddw}
  \chi_n({\bf q},i\omega_{n})&=&\frac{2\chi_{1}}{\chi_{2}+U\chi_{1}},
 \end{eqnarray}
 where
\begin{subequations}\label{xdw}
\begin{eqnarray}
\chi_{1}&=&\left|\begin{array}{cccc}
&\chi^{0}_{11}+\chi^{0}_{12} &2\chi^{0}_{14}U &2\chi^{0}_{13}U\\
&\chi^{0}_{14} &1+\chi^{0}_{34}U &-\chi^{0}_{12}U\\
&\chi^{0}_{13}&-\chi^{0}_{12}U &1+\chi^{0}_{43}U\\
\end{array}\right| \\
\chi_{2}&=&\left|\begin{array}{ccc}
&1+\chi^{0}_{34}U &-\chi^{0}_{12}U\\
&-\chi^{0}_{12}U &1+\chi^{0}_{43}U\\
\end{array}\right|.
\end{eqnarray}
\end{subequations}
According to the fluctuation-dissipation theory, the density dynamical structure factor $S({\bf q},{\omega})$ is connected to the imaginary part of the density response function $\chi_{n}$ by
\begin{eqnarray}\label{sqw}
  S({\bf q},{\omega})&=&-\frac{1}{\pi}{\rm Im}\chi_n\left({\bf q},i\omega_{n}\to \omega+i\delta\right),
 \end{eqnarray}
where ${\bf q}$ and $\omega$ are respectively the transferred momentum and energy. $\delta$ is a small positive number in numerical calculation (usually we set $\delta=0.003$).

\section{Results at half-filling}
\label{halfres}
\begin{figure*}[t]
\centering
\includegraphics[width=0.9\textwidth]{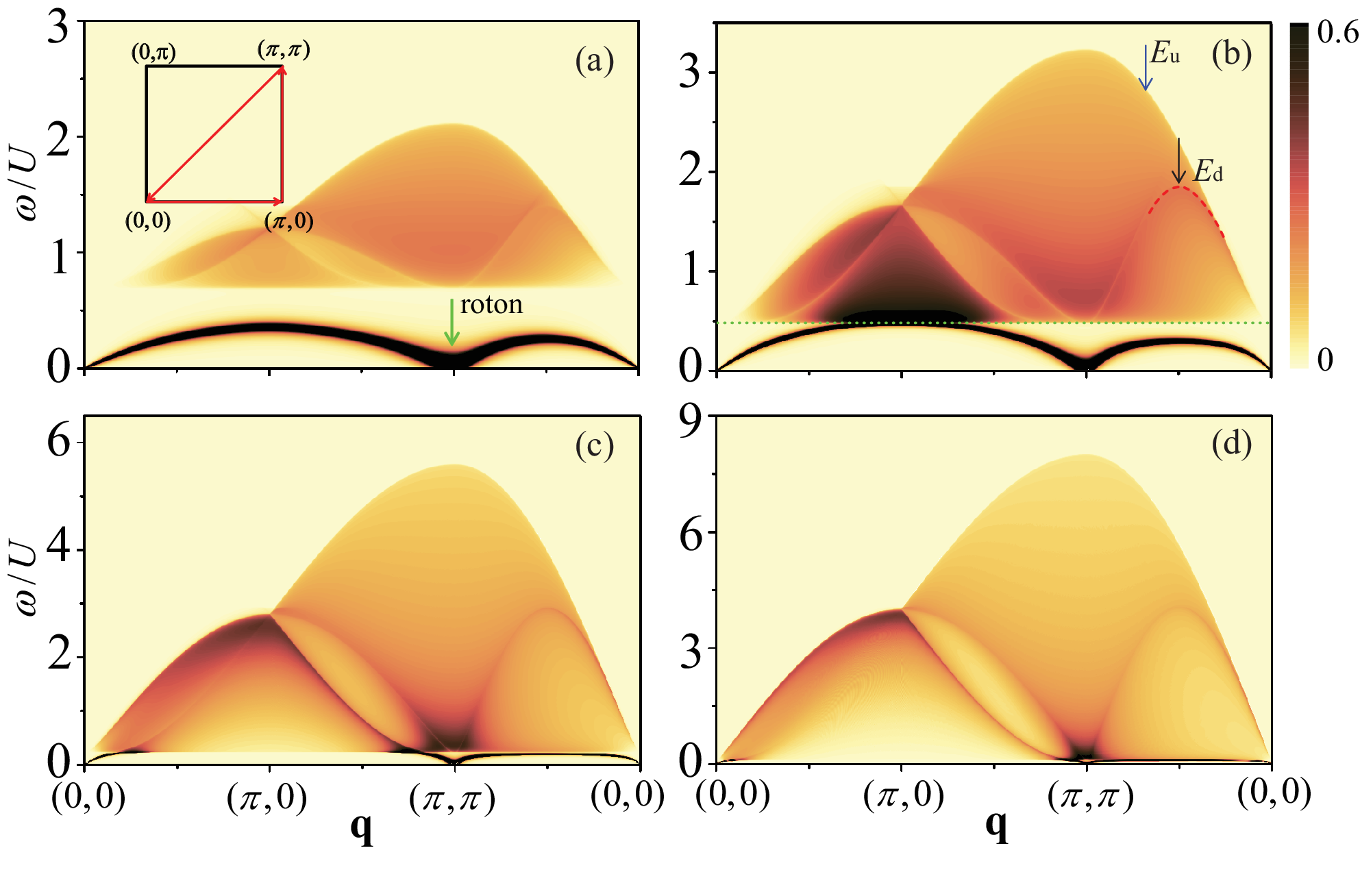}
\caption{\label{fig2} Color maps of the dynamical structure factor $S({\bf q},{\omega})$ as functions of transferred energy and momentum along the high-symmetry directions for hopping strength (a) $t/U=0.25$, (b) $0.4$, (c) $0.7$, and (d) $1.0$ at half-filling $n=1$. The red dashed line marks the dispersion of $E_{\rm d}$. The green dotted line indicates the minimum energy to break a Cooper pair. Inset in (a): a quarter of the first BZ and three high-symmetry directions (red arrows).}
\end{figure*}

 We first study the dynamical structure factor of 2D attractive Fermi-Hubbard model at half-filling ($n=1$). By analyzing the density dynamical structure factor $S({\bf q},{\omega})$ under different transferred momenta ${\bf q}$, both the collective excitations and single-particle excitations of Fermi atomic gases can be obtained. We calculate the energy and momentum dependencies of $S({\bf q},{\omega})$ and present its spectral weight distribution along the high-symmetry directions in the BZ for varying hopping strengths $t/U$, as shown in Fig. \ref{fig2}.
In the low-energy region, $S(\bf{q},\omega)$ displays sharp peaks, which denotes two gapless collective modes. The first mode emerges from ${\bf q}=[0,0]$ and increases almost linearly in the low-momentum region along both $[0,0]\rightarrow [\pi,0]$ and $[0,0]\rightarrow [\pi,\pi]$. This is the characteristic of the phonon mode originating from the spontaneously $U(1)$ symmetry breaking of pairing gap (or order parameter). The slope of the phonon mode at ${\bf q}=[0,0]$ defines the sound speed. The second mode is the roton mode appearing near ${\bf q}=[\pi,\pi]$. The roton mode can be understood through a global pseudospin $SU(2)$ symmetry breaking \cite{Zhang1990,Ganesh2009,Qin2022}. There is a degeneracy between superfluid and charge density wave (CDW). In this paper, we don't discuss the competition between them. As the momentum increases, the phonon mode gradually merges into the single-particle excitations, and shows a finite expansion width because of the scattering with the single-particle excitations. The appearance of two collective modes had been confirmed by the QMC simulations \cite{Vitali2020}.

In the high-energy regime, the excitations enter the single-particle continuum dominated by the pair-breaking effect. The energy $E_{\bf{k+q}}+E_{\bf{k}}$ required to break Cooper pairs at a certain transferred momentum ${\bf q}$ is determined by the quasiparticle spectrum $E_{\bf{k}}$. This pair-breaking process forms a continuous excitation band, and the minimum energy (threshold) min$[E_{\bf{k+q}}+E_{\bf{k}}]$ is labeled by a green dotted line in Fig. \ref{fig2}\textcolor{blue}{(b)}. This green horizontal line corresponds to the transferred energy $\omega=2\Delta$, marking the minimum pair-breaking threshold and providing an experimental strategy for measuring the pairing gap. However, this strategy to measure the pairing gap becomes unreliable in systems with complex band structures (e.g., under SOC), where the threshold varies with ${\bf q}$. It is clearly seen in Fig. \ref{fig2} that increasing $t$ makes the threshold move to the low-energy region since the pairing gap is suppressed. Furthermore, compared with the RPA results, the QMC results reveal that the upper single-particle excitation branch has a significantly lower energy, suggesting the shortcomings of the RPA theory in strong interacting systems. To address this, we introduce the quasiparticle coherent weight $Z_{F}$ \cite{Schafer2021,Hafermann2014,Feng2015,Anderson2004} and a normal-state full Green's function including many-body interaction. This Green's function $g({\bf k},\omega)=1/(\omega-\xi_{\bf k}-\Sigma({\bf k},\omega))$, where $\Sigma({\bf k},\omega)$ is the self-energy and can be decoupled as $\Sigma({\bf k},\omega)=\Sigma_{e}({\bf k},\omega)+\omega\Sigma_{o}({\bf k},\omega)$. The $Z_{F}$ is defined as: $Z^{-1}_{F}({\bf k},\omega)=1-\Sigma_{o}({\bf k},\omega)$. Under the static limit approximation, $Z_{F}=Z_{F}({\bf k=k_{F}},\omega=0)$ by taking the Fermi momentum ${\bf k=k_{F}}$, simplifying the full Green's function to $g({\bf k},\omega)=Z_{F}/(\omega-Z_{F}\xi_{\bf k})={Z_{F}}/(\omega-\bar{\xi}_{\bf k})$, where $\bar{\xi}_{\bf k}=Z_{F}\xi_{\bf k}$ is the renormalized energy spectrum. For the non-interacting systems, $Z_{F}=1$. Increasing the interaction strength reduces $Z_{F}$, lowering the upper branch energy. In a superfluid, $Z_{F}$ also suppresses the pairing gap, explaining the smaller pairing gap in QMC than that through the RPA. In this paper, we do not discuss the effect of $Z_{F}$.

The band width of the single-particle excitations $W$ increases with the hopping term, namely, $W=8t$.
 Along $[0,0]\rightarrow [\pi,\pi]$ direction, $S({\bf q},{\omega})$ exhibits a distinct upper boundary marked by the blue arrow in panel (b), and it is described as:
\begin{eqnarray}\label{upb}
 E_{\rm u}=W{\rm sin}\left({\bf q}/2\right)=8t{\rm sin}\left({\bf q}/2\right).
\end{eqnarray}
The lower contour marked by the black arrow in panel (b) is obtained as:
\begin{eqnarray}\label{downb}
 E_{\rm d}=1.83 {\rm sin}({\bf q}).
\end{eqnarray}
The red-dashed line shows the dispersion of $E_{\rm d}$. Physically, $E_{\rm d}$ is not a collective mode but analogous to the 1D lower boundary of $E_{\rm d}=4t{\sin}({\bf q})$ at $U=0$ \cite{Han2022,Nocera2016}. In the Heisenberg model, $E_{\rm d}$ corresponds to the des Cloizeaux-Pearson (dCP) dispersion \cite{Nocera2016}. Owing to the effect of pairing gap, here $E_{\rm d}=1.83{\sin}({\bf q})$. $E_{\rm d}$ as a function of ${\bf q}$ is characterized by a double periodicity compared with $E_{\rm u}$. Both $E_{\rm u}$ and $E_{\rm d}$ originate from the single-particle excitations which can be understood by: $\hbar\omega_{\bf kq}=\xi_{{\bf k}+{\bf q}}-\xi_{\bf k}$. Along $[0,0]\rightarrow [\pi,\pi]$, this becomes $\hbar\omega_{\bf kq}=8t\sin\left({\bf k}+{\bf q}/2\right)\sin\left({\bf q}/2\right)$. When $\sin\left({\bf k}+{\bf q}/2\right)=1$, the maximum excitation $\hbar\omega^{max}_{\bf kq}=8t\sin({\bf q}/2)=E_{\rm u}$. At $n=1$, the Fermi momentum $k_{F}=\pi/2$. When ${\bf k}=k_{F}$, $\hbar\omega_{\bf kq}=4t\sin({\bf q})=E_{\rm d}$.
\section{Results away from half filling}
\label{awhalf}
\begin{figure*}[t]
\centering
\includegraphics[width=0.9\textwidth]{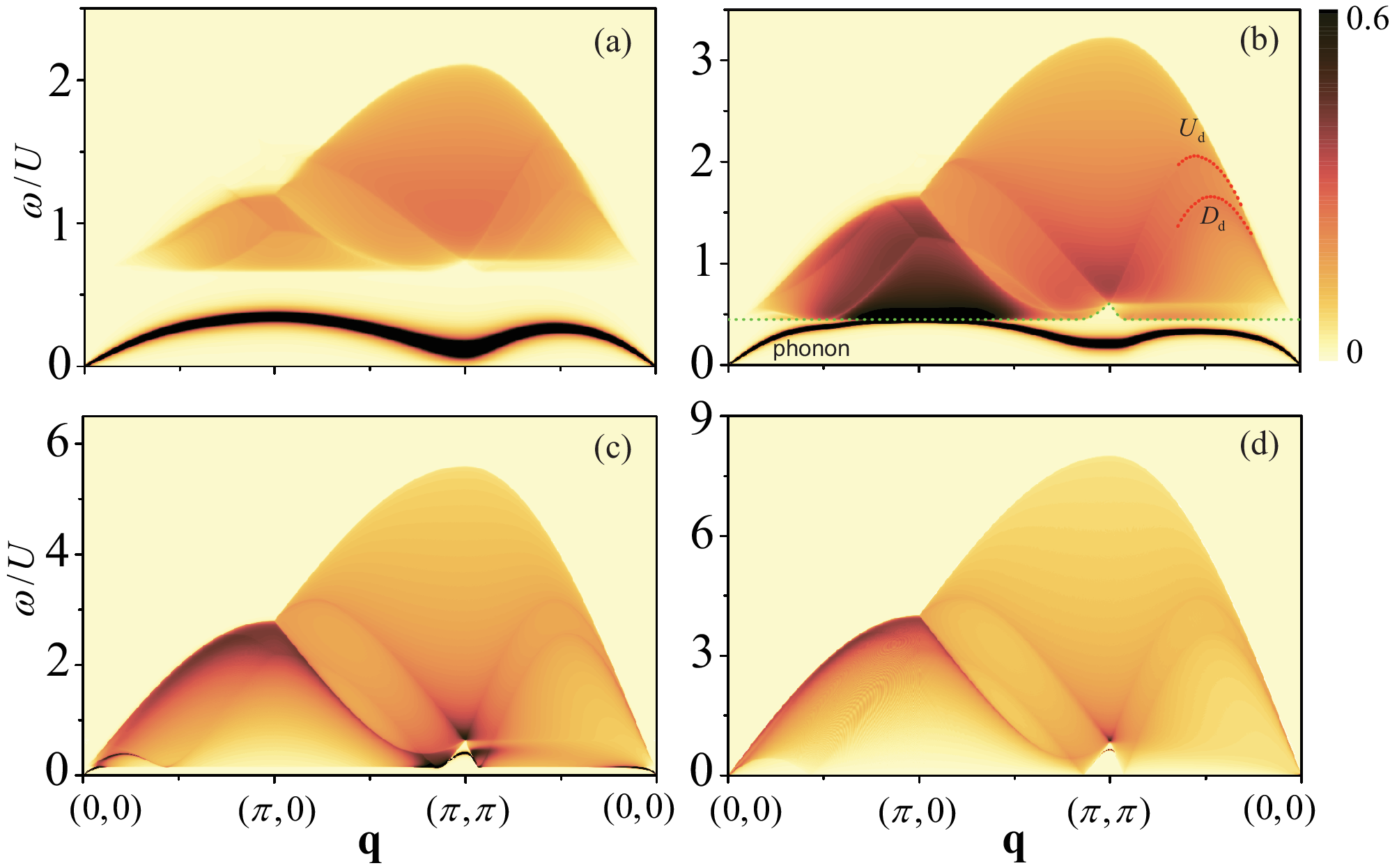}
\caption{\label{fig3} $S({\bf q},{\omega})$ along the high-symmetry directions for (a) $t/U=0.25$, (b) $0.4$, (c) $0.7$, and (d) $1.0$ at $n=0.8$. Panel (b) shows $E_{\rm d}$ splitting into $U_{\rm d}$ and $D_{\rm d}$ (blue dotted lines) induced by doping.
}
\end{figure*}

 Doping influences particle density and alters the Fermi energy, thereby greatly affecting the dynamical excitations. The particle density $n=1-\delta$, where $\delta$ is the doping concentration. We discuss the energy and momentum dependencies of $S({\bf q},{\omega})$ at $n=0.8$ (away from the half-filling). In Fig. \ref{fig3}, we plot the contour of $S({\bf q},{\omega})$ along the high-symmetry directions in the BZ for different hopping strengths. Three key differences appear when comparing $n=0.8$ with the half-filling, namely, the molecular excitations at ${\bf q}=[\pi, \pi]$, the split of $E_{\rm d}$, and the doping variation of sound speed. The following three subsections will elaborate on these differences

\subsection{Molecular excitations at ${\bf q}=[\pi, \pi]$ and its related pairing gap}
\begin{figure}[h!]
\includegraphics[scale=0.45]{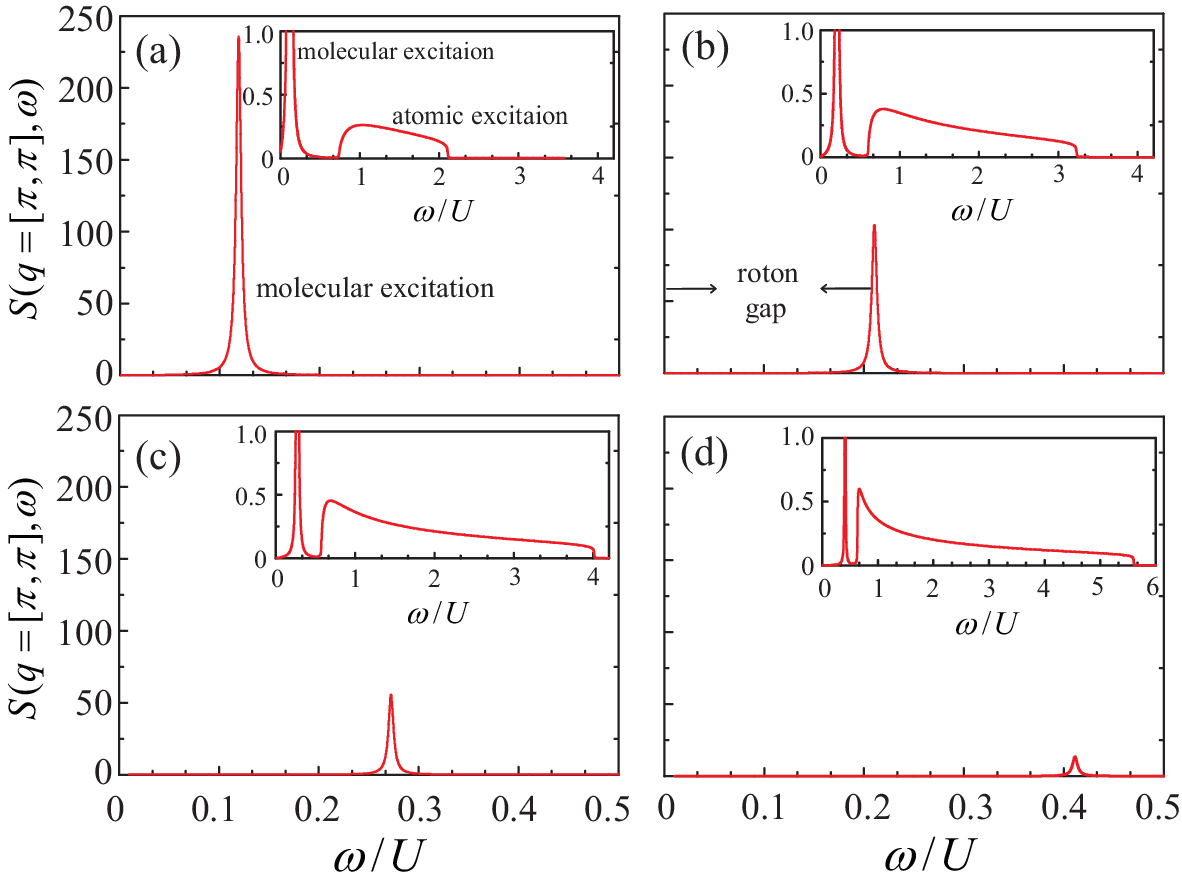}
\caption{ $S({\bf q}=[\pi, \pi],{\omega})$ as a function of  $\omega$ for the hopping strengths (a) $t/U=0.25$, (b) $0.4$, (c) $0.5$, and (d) $0.7$  at $n=0.8$. Insets magnify the atomic excitation regime.
\label{fig4}}
\end{figure}
Compared with Fig. \ref{fig2}, the minimum of the roton mode at $[\pi, \pi]$ moves above the zero energy at $n=0.8$. Thus, this roton mode is gapped which arises from the strong local density correlations \cite{Ganesh2009}. In optical lattices, this gapped roton mode is always well separated from single-particle excitations and is explained as the molecular Cooper-pair excitations here \cite{Zhang1990}. To show this clearly, $S({\bf q},{\omega})$ as a function of $\omega$ for different $t$ are shown in Fig. \ref{fig4}. Insets magnify the atomic excitation regime. Obviously, $S({\bf q},{\omega})$ consists of a sharp low-energy peak and a broad high-energy single-particle excitation band. This sharp peak corresponds to the excitation of the bosonic molecules from a molecular condensate while the broad single-particle excitation band is the result of atomic (particle-hole) excitations. As $t$ increases ($U$ decreases), the weight of the molecular peak decreases while the atomic excitation band increases quickly.

\begin{figure}[h!]
\includegraphics[scale=0.5]{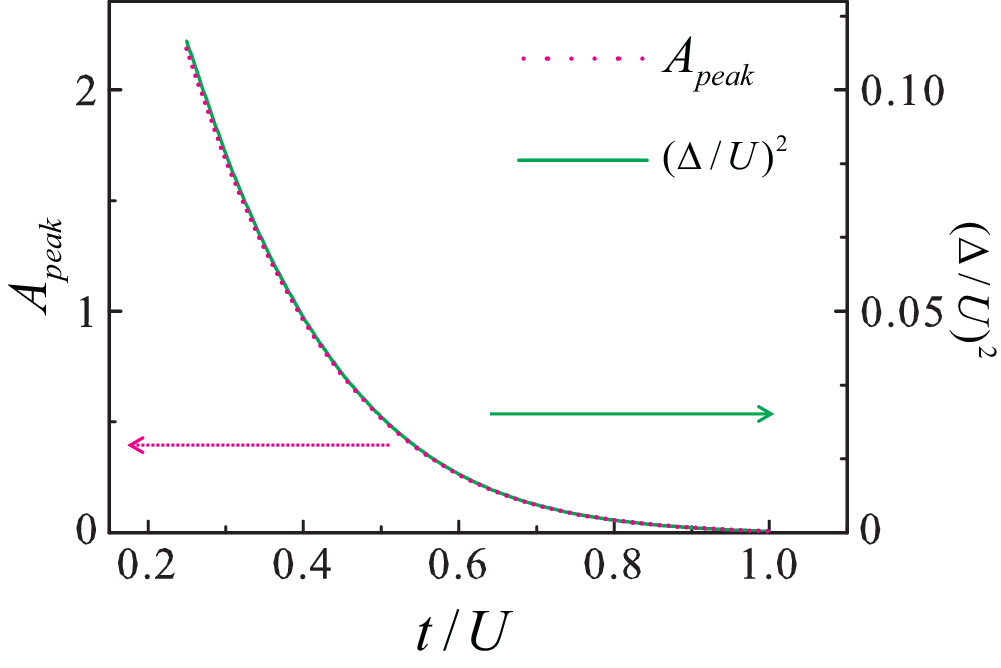}
\caption{ The area of the molecular peak $A_{peak}$ (pink dotted line) and the square of the pairing gap $\Delta^{2}$ (green solid line) as a function of $t$ at $n=0.8$. \label{fig5}}
\end{figure}
The weight of the molecular peak can be quantified by its spectral area. In Fig.~\ref{fig5}, we display the relation between the area of the molecular peak $A_{peak}$ (pink dotted line) and the hopping strength $t$, compared with the square of the pairing gap $\Delta^{2}$ (green solid line). Our results show that $A_{peak}$ has almost the same $t$ dependence as $\Delta^{2}$. Both quantities are particularly large at small $t$ and decrease gradually with increasing $t$ from the intermediate to the weak coupling regime, indicating that $\Delta^{2}$ governs the molecular peak at ${\bf q}=[\pi, \pi]$, suggesting $\Delta$ can be experimentally measured by detecting $S({\bf q}=[\pi, \pi],{\omega})$. It is worth noticing that this method to measure $\Delta$ is universal. For 3D cubic optical lattices, our calculations demonstrate that $A_{peak}$ at ${\bf q}=[\pi, \pi, \pi]$ (roton mode) remains proportional to $\Delta^{2}$. Thus, the method of the pairing gap measurement does not depend on the spatial dimension, demonstrating the dimensional universality. In particular, this method is also suitable for the SOC Fermi systems, where $\Delta$ is difficult to measure owing to the complex band structures \cite{Zhao2023}.

\subsection{The split of $E_{\rm d}$}
\begin{figure}[h!]
\includegraphics[scale=0.5]{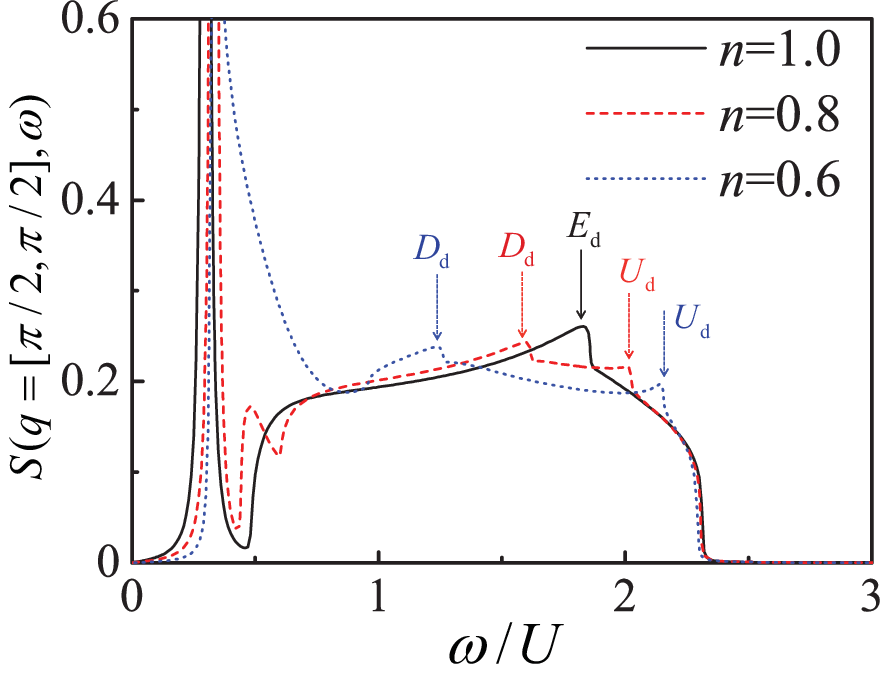}
\caption{ $S({\bf q}=[\pi/2,\pi/2],{\omega})$ as a function of $\omega$ for $n=1.0$ (black solid line), $n=0.8$ (red dashed line), and $n=0.6$ (blue dotted line) at $t/U=0.4$.
\label{fig6}}
\end{figure}
Doping also influences the atomic excitations. Compared with the half-filling case in Fig. \ref{fig2}, the boundary $E_{\rm d}$ splits due to doping. To show this clearly, we calculate $S({\bf q}=[\pi/2,\pi/2],{\omega})$ as a function of ${\omega}$ for different doping concentrations at $t/U=0.4$ in Fig. \ref{fig6}.

At the low-energy region, a sharp excitation peak appears as the signature of a collective mode. Another broad band at larger energy corresponds to the atomic excitations. At half-filling (red solid line) $n=1.0$, a characteristic peak at $\omega/U=1.83$ (marked by a black arrow) corresponds to $E_{\rm d}$. However, when $n=0.8$ and $0.6$, $E_{\rm d}$ splits into two branches: $U_{\rm d}$ and $D_{\rm d}$ (marked by red and blue dotted lines, respectively), which also exist in the normal state. This split of $E_{\rm d}$ is unrelated to the interaction term $H_{\rm int}=-\sum_{{\bf k}}(\Delta^{*}C_{{\bf k}\downarrow}C_{-{\bf k}\uparrow}+H.c.)$ in Eq. \ref{Humodel2}. Moreover, as doping increases, the gap between $U_{\rm d}$ and $D_{\rm d}$ increases. Along $[0,0]\rightarrow [\pi,\pi]$, the Fermi momentum $k_{F}=\pi/2$ at $n=1.0$. As $n$ decreases, $k_{F}$ decreases, thereby altering the single-particle excitation spectrum.

\subsection{Hopping dependence of sound speed}
\begin{figure}[h!]
\includegraphics[scale=0.4]{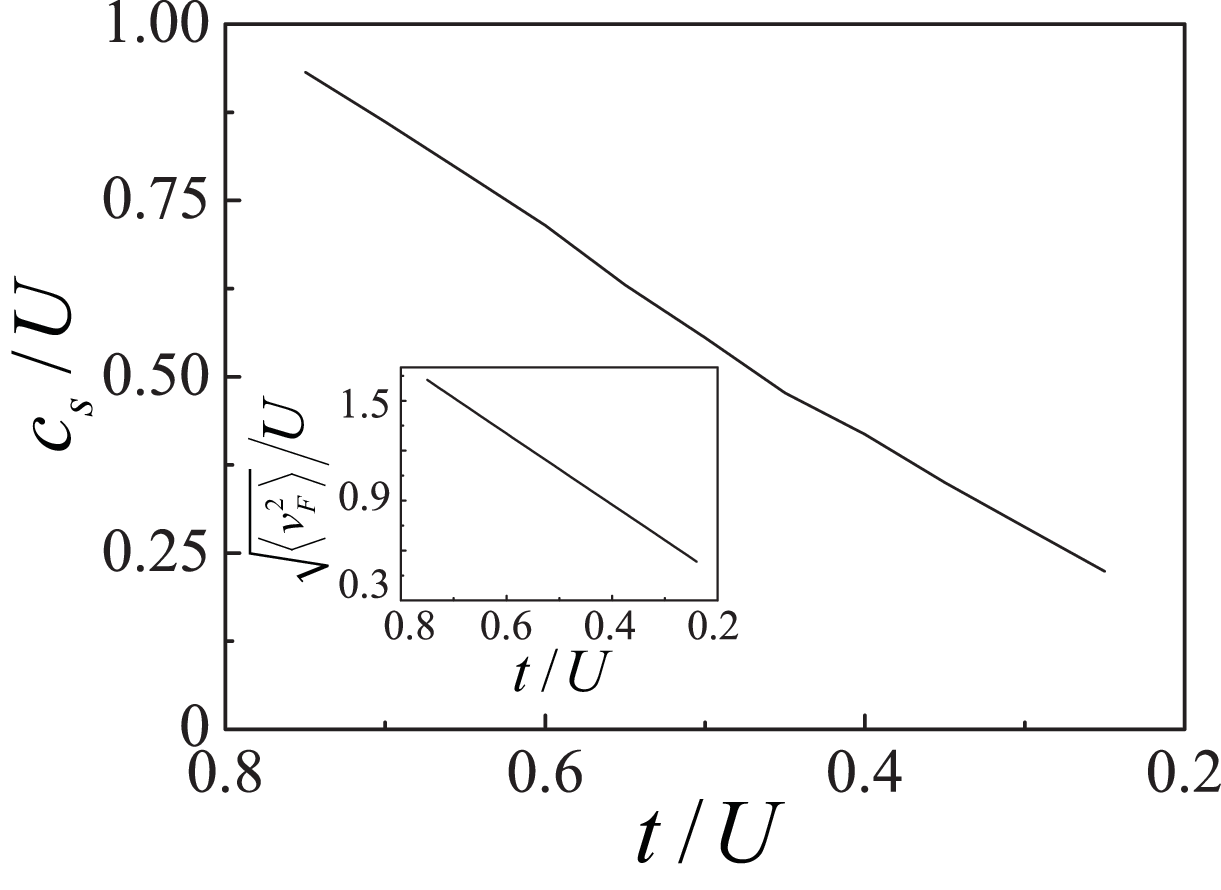}
\caption{Sound speed $c_{s}$ as a function of $t$ at $n=0.8$. Inset: $\sqrt{\left<v^{2}_{F}\right>}$ versus $t$.
\label{fig7}}
\end{figure}
 The slope of the Goldstone phonon mode in the limit $\omega \rightarrow 0$ and ${\bf q} \rightarrow 0$ defines the sound speed, $c_{\rm s}=\omega/|{\bf q}|$. The sound speed depends on the interaction strength parameterized by $t/U$. The sound speed $c_{s}$ as a function of $t$ is plotted in Fig. \ref{fig7}. Our theoretical results demonstrate that $c_{s}$ decreases with decreasing $t$ from the intermediate to the weak coupling regimes, qualitatively consistent with $^{6}{\rm Li}$ Fermi gas experiments \cite{Hoinka17,Sobirey2022}. The qualitatively behavior of $c_{s}$ can be explained by the squared Fermi velocity $\sqrt{\left<v^{2}_{F}\right>}$, which will be introduced in Fig. \ref{fig9}.

\section{Doping dependence of the dynamical structure factor}
\label{Dopingdsf}
\begin{figure}[h!]
\includegraphics[scale=0.42]{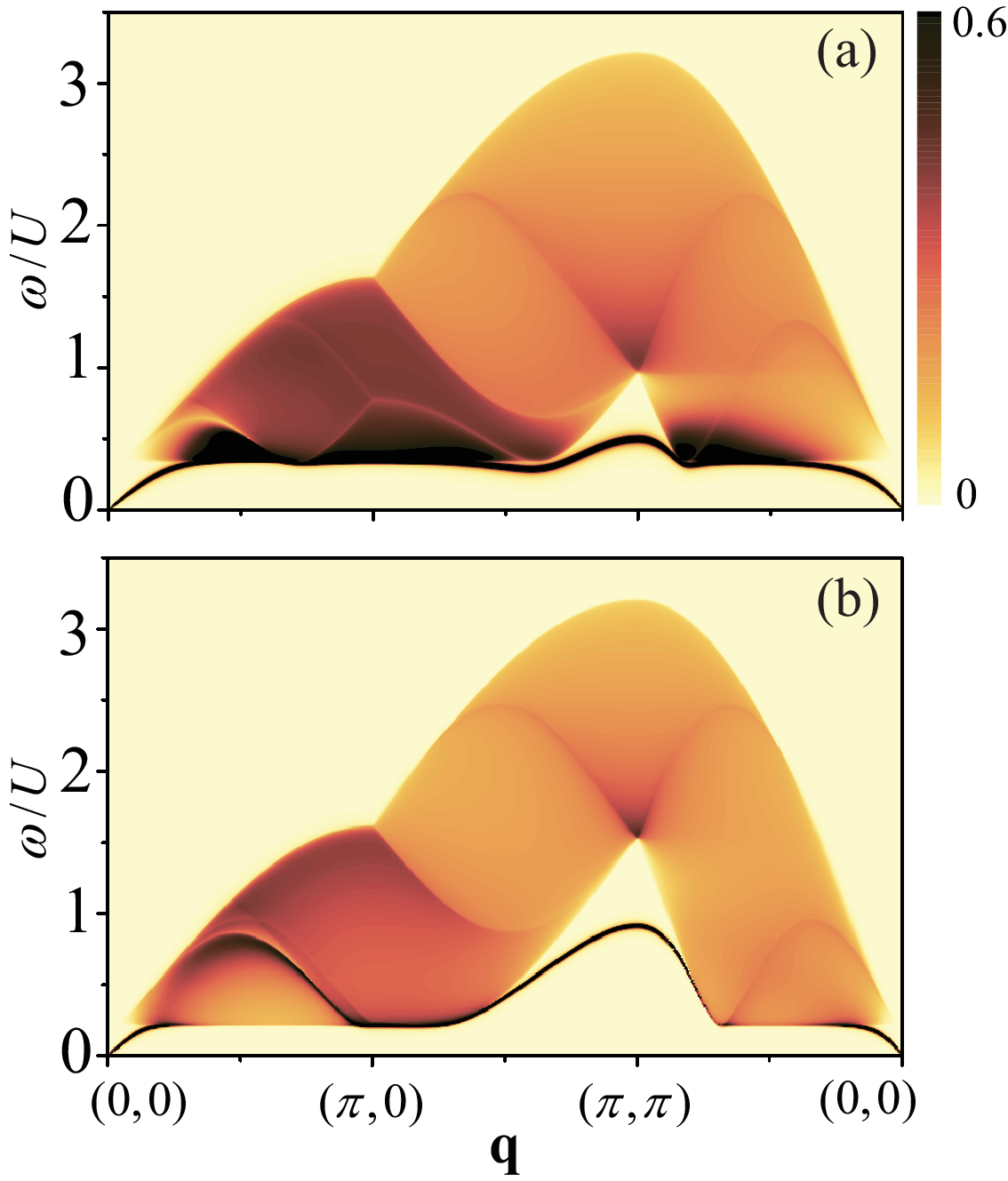}
\caption{Color maps of $S({\bf q},{\omega})$ for (a) $n=0.6$ and (b) $n=0.4$ with $t/U=0.4$.
\label{fig8}}
\end{figure}

\begin{figure}[h!]
\includegraphics[scale=0.4]{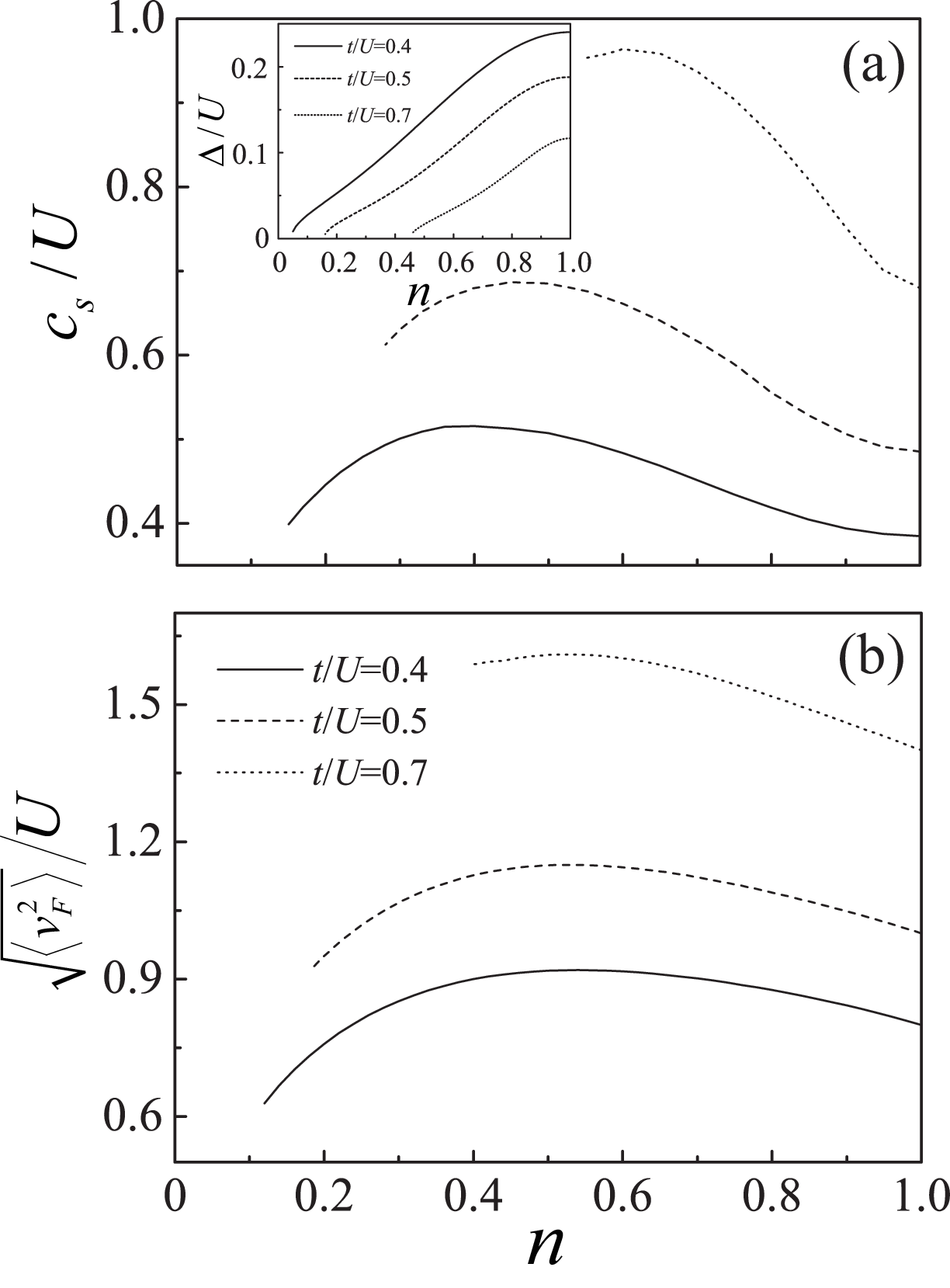}
\caption{(a) $c_{s}$ and (b) $\sqrt{\left<v^{2}_{F}\right>}$ as functions of $n$ for $t/U=0.4$ (solid line), $0.5$ (dashed line), and $0.7$ (dotted line). Inset to (a):  the corresponding $\Delta$ as a function of $n$.
\label{fig9}}
\end{figure}

Here we discuss the relation between dynamical excitations and the doping concentration. In Fig. \ref{fig8}, we plot $S({\bf q},{\omega})$ along the high-symmetry directions of the BZ for (a) $n=0.6$ and (b) $n=0.4$ with $t/U=0.4$.
Our results show that the molecular peak at ${\bf q}=[\pi,\pi]$ moves to the higher energy, creating an enlarged roton gap. In particular, the sound speed exhibits doping dependence. To clarify this, we plot $c_{s}$ as a function of $n$ in Fig. \ref{fig9}\textcolor{blue}{(a)}.
Our results show that the sound speed first increases and then decreases as $n$ increases. This behavior can be qualitatively understood through a weak interaction theory \cite{Belkhir1994}. The sound speed is closely related to the squared Fermi velocity, $c_{s}=\sqrt{\left<v^{2}_{F}\right>[1-UN(0)]/2}$, where $N(0)$ is the density of states (DOS) at the Fermi energy, is given by $N(0)=(2\pi)^{-2}\int{d^{2}{\bf k} \delta(\xi_{{\bf k}})}$. Therefore, $N(0)$ is proportional to the Fermi surface length or $n$. The Fermi velocity $v_{F}={\partial \xi_{\bf k}}/{\partial {\bf k}}|_{{\bf k}=[k_{Fx},k_{Fy}]}$ is evaluated at the Fermi wave vector determined by $\xi_{{\bf k}=[k_{Fx},k_{Fy}]}=0$ and the self-consistent equations Eq. \ref{twoequtions}. Thus $k_{F}$, $\mu$ and $\Delta$ can be obtained self-consistently for a given $n$. We define $\left<v^{2}_{F}\right>=\frac{1}{N_{0}}\sum_{\bf k_{F}}v^{2}_{F}$ with $N_{0}=120$ selected points along the Fermi surface.  We plot $\sqrt{\left<v^{2}_{F}\right>}$ as a function of $n$ in Fig. \ref{fig9}\textcolor{blue}{(b)}. It is shown that $\sqrt{\left<v^{2}_{F}\right>}$ has qualitatively the same $n$ dependence as that of $c_{s}$. At half-filling, $v_{F}$ is anisotropic in the BZ. When $n$ is small, the Fermi surface shrinks, reducing the $v_{F}$ anisotropy. In this case, the physical properties of an optical lattice approximate the continuum Fermi gases. In the low-momentum region, the cosine function can be expanded as: $\cos{k}=1-k^{2}/2$, yielding that the energy spectrum of the optical lattice can be approximated as: $\xi_{\bf k}=Zt(k^{2}_{x}+k^{2}_{y})-tZ-\mu$, which matches the continuum case $\xi_{\bf k}=(k^{2}_{x}+k^{2}_{y})/2m-\mu$. Moreover, as $n\rightarrow 0$, $N(0)\rightarrow 0$, yielding $c_{s}=\sqrt{\left<v^{2}_{F}\right>/2}$, consistent with ideal continuum Fermi gases.

\section{Summary}
\label{summary}
In conclusion, the doping and hopping dependencies of the dynamical structure factor in the 2D attractive Fermi-Hubbard model were studied within RPA theory. Two collective modes emerge: the phonon mode at small transferred momenta and the roton mode at transferred momentum ${\bf q}=[\pi,\pi]$ regime. The roton mode corresponds to Cooper pair molecular excitations. First, the molecular excitation peak area at ${\bf q}=[\pi,\pi]$ scales with the square of the pairing gap at fixed doping. This motivates a universal the pairing gap measurement strategy, being applicable to 2D/3D and SOC optical lattices. While this result is obtained from the weak coupling to the intermediate coupling regime within the RPA, we conjecture that it is always valid even in strong interactions. Because this strategy does not depend on dimension and SOC interaction. And the relation between the roton mode the pairing gap is also confirm by both Zhang's theory and Dyke{\it et al.}'s Bragg experiments. It is also essential to verify this strategy beyond the RPA. Second, the atomic excitation peak splits into two branches when the system is away from half-filling, and the splitting magnitude increases with increasing doping. Third, the sound speed at a given doping is suppressed by interaction strength and governed by the squared Fermi velocity.

\section{Acknowledgements}
This work was supported by the funds from the National
Natural Science Foundation of China under Grant No.11547034 (H.Z.), Grants No. U23A2073 (P.Z.)

Perceptually uniform color maps ('lajolla') are used in this study \cite{Crameri2018}.
\section{Appendix}
 The mean-field response function $\chi^0$ of 2D interacting Fermi atoms in a square optical lattice is numerically calculated, and all 6 independent matrices elements of $\chi^0$ are
\begin{eqnarray}\label{a1}
 \chi^{0}_{11}&=&\frac{1}{4}\sum_{\bf k} \left[1+\frac{\xi_{{\bf k}}\xi_{{\bf k}+{\bf q}}}{E_{{\bf k}}E_{{\bf k}+{\bf q}}}\right]F^{(1)}_{{\bf k},{\bf q}} \notag\\
 &+&\frac{1}{4}\sum_{\bf k} \left[1-\frac{\xi_{{\bf k}}\xi_{{\bf k}+{\bf q}}}{E_{\bf k}E_{{\bf k}+{\bf q}}}\right]F^{(2)}_{{\bf k},{\bf q}},\nonumber
 \end{eqnarray}
\begin{eqnarray}\label{b}
 \chi^{0}_{12} =-\frac{1}{4}\sum_{\bf k} \frac{\Delta^{2}}{E_{\bf k}E_{{\bf k}+{\bf q}}} \left[F^{(1)}_{{\bf k},{\bf q}}-F^{(2)}_{{\bf k},{\bf q}}\right] ,\nonumber
 \end{eqnarray}
\begin{eqnarray}\label{ct}
 \chi^{0}_{13}=&-&\frac{\Delta}{4}\sum_{\bf k} \frac{\xi_{\bf k}+\xi_{{\bf k}+{\bf q}}}{2E_{\bf k}E_{{\bf k}+{\bf q}}} \left[F^{(1)}_{{\bf k},{\bf q}}-F^{(2)}_{{\bf k},{\bf q}}\right] \notag\\
 &+&\frac{\Delta}{4}\sum_{\bf k} \frac{E_{{\bf k}+{\bf q}}-E_{\bf k}}{2E_{\bf k}E_{{\bf k}+{\bf q}}} F^{(3)}_{{\bf k},{\bf q}}\notag \\
 &-&\frac{\Delta}{4}\sum_{\bf k} \frac{E_{{\bf k}+{\bf q}}+E_{\bf k}}{2E_{\bf k}E_{{\bf k}+{\bf q}}}F^{(4)}_{{\bf k},{\bf q}}  ,\notag
 \end{eqnarray}
\begin{eqnarray}\label{c}
 \chi^{0}_{14}=&-&\frac{\Delta}{4}\sum_{\bf k} \frac{\xi_{\bf k}+\xi_{{\bf k}+{\bf q}}}{2E_{\bf k}E_{{\bf k}+{\bf q}}} \left[F^{(1)}_{{\bf k},{\bf q}}-F^{(2)}_{{\bf k},{\bf q}}\right] \notag\\
 &-&\frac{\Delta}{4}\sum_{\bf k} \frac{E_{{\bf k}+{\bf q}}-E_{\bf k}}{2E_{\bf k}E_{{\bf k}+{\bf q}}} F^{(3)}_{{\bf k},{\bf q}}\notag \\
 &+&\frac{\Delta}{4}\sum_{\bf k} \frac{E_{{\bf k}+{\bf q}}+E_{\bf k}}{2E_{\bf k}E_{{\bf k}+{\bf q}}}F^{(4)}_{{\bf k},{\bf q}}  ,\notag
 \end{eqnarray}
\begin{eqnarray}\label{h}
 \chi^{0}_{43}
 &=&\frac{1}{4}\sum_{\bf k} \left[ 1-\frac{\xi_{{\bf k}}\xi_{{\bf k}+{\bf q}}}{E_{\bf k}E_{{\bf k}+{\bf q}}} \right]F^{(1)}_{{\bf k},{\bf q}} \notag \\
 &+&\frac{1}{4}\sum_{\bf k} \left[ 1+\frac{\xi_{{\bf k}}\xi_{{\bf k}+{\bf q}}}{E_{{\bf k}}E_{{\bf k}+{\bf q}}} \right]F^{(2)}_{{\bf k},{\bf q}} \notag \\
 &+&\frac{1}{4}\sum_{\bf k} \left[ \frac{\xi_{\bf k}}{E_{\bf k}}-\frac{\xi_{{\bf k}+{\bf q}}}{E_{{\bf k}+{\bf q}}} \right]F^{(3)}_{{\bf k},{\bf q}} \notag \\
 &-&\frac{1}{4}\sum_{\bf k} \left[ \frac{\xi_{\bf k}}{E_{\bf k}}+\frac{\xi_{{\bf k}+{\bf q}}}{E_{{\bf k}+{\bf q}}} \right]F^{(4)}_{{\bf k},{\bf q}} ,\nonumber
\end{eqnarray}
\begin{eqnarray}\label{ht}
 \chi^{0}_{34}
 &=&\frac{1}{4}\sum_{\bf k} \left[ 1-\frac{\xi_{{\bf k}}\xi_{{\bf k}+{\bf q}}}{E_{\bf k}E_{{\bf k}+{\bf q}}} \right]F^{(1)}_{{\bf k},{\bf q}} \notag \\
 &+&\frac{1}{4}\sum_{\bf k} \left[ 1+\frac{\xi_{{\bf k}}\xi_{{\bf k}+{\bf q}}}{E_{{\bf k}}E_{{\bf k}+{\bf q}}} \right]F^{(2)}_{{\bf k},{\bf q}} \notag \\
 &-&\frac{1}{4}\sum_{\bf k} \left[ \frac{\xi_{\bf k}}{E_{\bf k}}-\frac{\xi_{{\bf k}+{\bf q}}}{E_{{\bf k}+{\bf q}}} \right]F^{(3)}_{{\bf k},{\bf q}} \notag \\
 &+&\frac{1}{4}\sum_{\bf k} \left[ \frac{\xi_{\bf k}}{E_{\bf k}}+\frac{\xi_{{\bf k}+{\bf q}}}{E_{{\bf k}+{\bf q}}} \right]F^{(4)}_{{\bf k},{\bf q}} .\nonumber
\end{eqnarray}
 The corresponding functions in above equations $F^{(1)}_{{\bf k},{\bf q}}$, $F^{(2)}_{{\bf k},{\bf q}}$, $F^{(3)}_{{\bf k},{\bf q}}$, $F^{(4)}_{{\bf k},{\bf q}}$ are defined as
\begin{eqnarray}\label{Fkq}
 F^{(1)}_{{\bf k},{\bf q}}&=&A({\bf k},{\bf q},i\omega_{n})-B({\bf k},{\bf q},i\omega_{n}),\nonumber\\
 F^{(2)}_{{\bf k},{\bf q}}&=&C({\bf k},{\bf q},i\omega_{n})-D({\bf k},{\bf q},i\omega_{n}),\nonumber\\
 F^{(3)}_{{\bf k},{\bf q}}&=&A({\bf k},{\bf q},i\omega_{n})+B({\bf k},{\bf q},i\omega_{n}),\nonumber\\
 F^{(4)}_{{\bf k},{\bf q}}&=&C({\bf k},{\bf q},i\omega_{n})+D({\bf k},{\bf q},i\omega_{n}),
\end{eqnarray}
where
\begin{eqnarray}\label{Akq}
A({\bf k},{\bf q},i\omega_{n})&=&\frac{f(E_{\bf k})-f(E_{{\bf k}+{\bf q}})}{i\omega_{n}+(E_{\bf k}-E_{{\bf k}+{\bf q}})}\nonumber\\
B({\bf k},{\bf q},i\omega_{n})&=&\frac{f(E_{\bf k})-f(E_{{\bf k}+{\bf q}})}{i\omega_{n}-(E_{\bf k}-E_{{\bf k}+{\bf q}})}\nonumber\\
C({\bf k},{\bf q},i\omega_{n})&=&\frac{1-f(E_{\bf k})-f(E_{{\bf k}+{\bf q}})}{i\omega_{n}-(E_{\bf k}+E_{{\bf k}+{\bf q}})}\nonumber\\
D({\bf k},{\bf q},i\omega_{n})&=&\frac{1-f(E_{\bf k})-f(E_{{\bf k}+{\bf q}})}{i\omega_{n}+(E_{\bf k}+E_{{\bf k}+{\bf q}})},
\end{eqnarray}
$f(E_{\bf k})$ and $f(E_{{\bf k}+{\bf q}})$ are Fermi distributions.


\begin{thebibliography}{00}
\makeatletter
\bibitem{Feld11} M. Feld, B. Fr{\"o}hlich, E. Vogt, M. Koschorreck and M. K{\"o}hl, \textit{Observation of a pairing pseudogap in a two-dimensional Fermi gas}, \href{https://doi.org/10.1038/nature10627}{Nature \textbf{480}, 75 (2011)}.
\bibitem{Stewart2008} J. T. Stewart, J. P. Gaebler and D. S. Jin, \textit{Using photoemission spectroscopy to probe a strongly interacting Fermi gas}, \href{https://doi.org/10.1038/nature07172}{Nature
   \textbf{454}, 744 (2008)}.
\bibitem{Frohlich2011} B. Fr{\"o}hlich, M. Feld, E. Vogt, M. Koschorreck, W. Zwerger and M. K{\"o}hl, \textit{Radio-Frequency Spectroscopy of a Strongly Interacting Two-Dimensional Fermi Gas}, \href{https://doi.org/10.1103/PhysRevLett.106.105301}{Phys. Rev. Lett. \textbf{106}, 105301 (2011)}.
\bibitem{Chin2004} C. Chin, M. Bartenstein, A. Altmeyer, S. Riedl, S. Jochim, J. Hecker Denschlag, and R. Grimm, \textit{Observation of the Pairing Gap in
a Strongly Interacting Fermi Gas}, \href{https://www.science.org/doi/10.1126/science.1100818}{Science \textbf{305}, 1128 (2004)}.
\bibitem{Sommer2012} A. T. Sommer, L. W. Cheuk, M. J. H. Ku, W. S. Bakr, and M. W. Zwierlein, \textit{Evolution of Fermion Pairing from Three to Two Dimensions}, \href{https://doi.org/10.1103/PhysRevLett.108.045302}{Phys. Rev. Lett. \textbf{108}, 045302 (2012)}.
\bibitem{Zhai2015} H. Zhai, \textit{Degenerate quantum gases with spin-orbit coupling: a review}, \href{https://iopscience.iop.org/article/10.1088/0034-4885/78/2/026001}{Rep. Prog. Phys. \textbf{78}, 026001 (2015)}.
\bibitem{Cheuk2012} L. W. Cheuk, A. T. Sommer, Z. Hadzibabic, T. Yefsah, W. S. Bakr and M. W. Zwierlein, \textit{Spin-Injection on Spectroscopy of a Spin-Orbit Coupled Fermi gas}, \href{https://doi.org/10.1103/PhysRevLett.109.095302}{Phys. Rev. Lett. \textbf{109}, 095302 (2012)}.
\bibitem{Wang2012} P. Wang, Z.-Q. Yu, Z. Fu, J. Miao, L. Huang, S. Chai, H. Zhai and J. Zhang, \textit{Spin-Orbit Coupled Degenerate Fermi Gases}, \href{https://doi.org/10.1103/PhysRevLett.109.095301}{Phys. Rev. Lett. \textbf{109}, 095301 (2012)}.
\bibitem{Wang2021} Z.-Y Wang, X.-C. Cheng, B.-Z. Wang, J.-Y. Zhang, Y.-H. Lu, C.-R. Yi, S. Niu, Y. Deng, X.-J. Liu, S. Chen, and J.-W. Pan, \textit{Realization of an ideal Weyl semimetal band in a quantum gas with 3D Spin-Orbit coupling}, \href{https://www.science.org/doi/10.1126/science.abc0105}{Science \textbf{372}, 271 (2021)}.
\bibitem{Wu2013} F. Wu, G.-C. Guo, W. Zhang, and W. Yi, \textit{Unconventional Superfluid in a Two-Dimensional Fermi gas with Anisotropic Spin-Orbit Coupling and Zeeman fields}, \href{https://doi.org/10.1103/PhysRevLett.110.110401}{Phys. Rev. Lett. \textbf{110}, 110401 (2013)}.
\bibitem{Han2023} R. Han, F. Yuan and H. Zhao, \textit{Phase diagram, band structure and density of states in
two-dimensional attractive Fermi-Hubbard model with Rashba spin-orbit coupling}, \href{https://doi.org/10.1088/1367-2630/acb80d}{{New J. Phys.} {\bf 25}, 023011 (2023)}.
\bibitem{Veeravalli08} G. Veeravalli, E. Kuhnle, P. Dyke, and C. J. Vale, \textit{Bragg Spectroscopy of a Strongly Interacting Fermi Gas}, \href{https://doi.org/10.1103/PhysRevLett.101.250403}{{Phys. Rev. Lett.} \textbf{101}, 250403 (2008)}.
\bibitem{Hoinka17} S. Hoinka, P. Dyke, M. G. Lingham, J. J. Kinnunen, G. M. Bruun and C. J. Vale, \textit{Goldstone mode and pair-breaking excitations
in atomic Fermi superfluid}, \href{https://doi.org/10.1038/nphys4187}{Nat. Phys. \textbf{13}, 943 (2017)}.
\bibitem{Dyke2023} P. Dyke, S. Musolino, H. Kurkjian, D. J. M. Ahmed-Braun, A. Pennings, I. Herrera, S. Hoinka, S. J. J. M. F. Kokkelmans, V. E. Colussi, C. J. Vale, \textit{Higgs oscillations in a unitary Fermi superfluid}, \href{https://doi.org/10.1103/PhysRevLett.132.223402}{Phys. Rev. Lett. \textbf{132}, 223402 (2024)}.
\bibitem{Biss2022} H. Biss, L. Sobirey, N. Luick, M. Bohlen, J. J. Kinnunen, G. M. Bruun, T. Lompe, and H. Moritz, \textit{Excitation Spectrum and Superfluid Gap of an Ultracold Fermi Gas}, \href{https://doi.org/10.1103/PhysRevLett.128.100401}{Phys. Rev. Lett. \textbf{128}, 100401 (2022)}.
\bibitem{Senaratne2022}	R. Senaratne, D. Cavazos-Cavazos, S. Wang, F. He, Y.-T. Chang, A. Kafle, H. Pu, X.-W. Guan, and R. G. Hulet, \textit{Spin-charge separation in a 1D Fermi gas with tunable interactions}, \href{https://www.science.org/doi/10.1126/science.abn1719}{Science \textbf{376}, 1305 (2022)}.
\bibitem{Li2022} X. Li, X. Luo, S. Wang, K. Xie, X. P. Liu, H. Hu, Y.-A. Chen, X.-C. Yao and J. W. Pan, \textit{Second sound attenuation near quantum criticality}, \href{https://www.science.org/doi/10.1126/science.abi4480}{Science, \textbf{375}, 528 (2022)}.
\bibitem{Pagano2014} G. Pagano, M. Mancini, G. Cappellini, P. Lombardi, F. Sch{\"o}fer, H. Hu, X.-J. Liu, J. Catani, C. Sias, M. Inguscio and L. Fallani, \textit{A one-dimensional liquid of fermions with tunable spin}, \href{https://doi.org/10.1038/nphys2878}{Nat. Phys. \textbf{10}, 198 (2014)}.
\bibitem{Combescot06} R. Combescot, S. Giorgini and  S. Stringari, \textit{Molecular signatures in the structure factor of an interacting Fermi gas}, \href{https://iopscience.iop.org/article/10.1209/epl/i2006-10165-x}{Europhys. Lett. \textbf{75}, 695 (2006)}.
\bibitem{Combescot2006} R. Combescot, M. Yu. Kagan, and S. Stringari, \textit{Collective mode of homogeneous superfluid Fermi gases in the BEC-BCS crossover}, \href{https://doi.org/10.1103/PhysRevA.74.042717}{Phys. Rev. A \textbf{74}, 042717 (2006)}.
\bibitem{Zou10} P. Zou, E. D. Kuhnle, C. J. Vale, and H. Hu, \textit{Quantitative
comparison between theoretical predictions and experimental results
for Bragg spectroscopy of a strongly interacting Fermi superfluid},
\href{https://doi.org/10.1103/PhysRevA.82.061605}{Phys. Rev. A \textbf{82}, 061605(R) (2010)}.
\bibitem{Zou16} P. Zou, F. Dalfovo, R. Sharma, X. J. Liu and H. Hu,  \textit{Dynamic structure factor of a strongly correlated Fermi superfluid
within a density functional theory approach}, \href{http://dx.doi.org/10.1088/1367-2630/18/11/113044}{{New J. Phys.} {\textbf 18}, 113044 (2016)}.
\bibitem{Zou18} P. Zou, H. Hu, and X.-J. Liu, \textit{Low-momentum dynamic structure factor of a strongly interacting Fermi gas at finite
temperature: The Goldstone phonon and its Landau damping}, \href{https://doi.org/10.1103/PhysRevA.98.011602}{Phys. Rev. A \textbf{98}, 011602(R) (2018)}.
\bibitem{Hu18} H. Hu, P. Zou, and X.-J. Liu, \textit{Low-momentum dynamic structure factor of a strongly interacting Fermi gas at finite temperature: A two-fluid hydrodynamic description}, \href{https://doi.org/10.1103/PhysRevA.97.023615}{Phys. Rev. A  \textbf{97}, 023615 (2018)}.
\bibitem{Zou2021} P. Zou, H. Zhao, L. He, X.-J. Liu, and H. Hu, \textit{Dynamic structure factors of a strongly interacting Fermi superfluid near an orbital Feshbach resonance across the phase transition from BCS to Sarma superfluid}, \href{https://doi.org/10.1103/PhysRevA.103.053310}{Phys. Rev. A \textbf{103}, 053310 (2021)}.
\bibitem{Kuhnle10} E. D. Kuhnle, H. Hu, X.-J. Liu, P. Dyke, M. Mark, P. D. Drummond, P. Hannaford, and C. J. Vale, \textit{Universal Behavior of Pair Correlations in a Strongly Interacting Fermi Gas}, \href{https://doi.org/10.1103/PhysRevLett.105.070402}{{Phys. Rev. Lett.} \textbf{105}, 070402 (2010)}.
\bibitem{Watabe10} S. Watabe, and T. Nikuni, \textit{Dynamic structure factor of the normal Fermi gas from the collisionless to the hydrodynamic regime}, \href{https://doi.org/10.1103/PhysRevA.82.033622}{Phys. Rev. A \textbf{82}, 033622 (2010)}.

\bibitem{Sobirey2022} L. Sobirey, H. Biss, N. Luick, M. Bohlen, H. Moritz, and T. Lompe, \textit{Observing the Influence of Reduced Dimensionality on Fermionic Superfluids}, \href{https://doi.org/10.1103/PhysRevLett.129.083601}{{Phys. Rev. Lett.} \textbf{129}, 083601 (2022)}.
\bibitem{Vitali17} E. Vitali, H. Shi, M. Qin, and S. Zhang, \textit{Visualizing the BEC-BCS crossover in a two-dimensional Fermi gas:Pairing gaps and dynamical response functions from ab initio computations}, \href{https://doi.org/10.1103/PhysRevA.96.061601}{Phys. Rev. A \textbf{96}, 061601(R) (2017)}.
\bibitem{Zhao2020} H. Zhao, X. Gao, W. Liang, P. Zou and F. Yuan, \textit{Dynamical structure factors of a two-dimensional Fermi superfluid within random phase approximation}, \href{https://doi.org/10.1088/1367-2630/abab3d}{{New J. Phys.} {\textbf 22}, 093012 (2020)}.
\bibitem{Gao2023} Z. Gao, L. He, H. Zhao, S.-G. Peng, and P. Zou, \textit{ Dynamic structure factor of one-dimensional Fermi superfluid with
spin-orbit coupling}, \href{https://doi.org/10.1103/PhysRevA.107.013304}{Phys. Rev. A \textbf{107}, 013304 (2023)}.

\bibitem{Bloch2008} I. Bloch, J. Dalibard and W. Zwerger, \textit{Many-body physics with ultracold gases}, \href{https://doi.org/10.1103/RevModPhys.80.885}{Rev. Mod. Phys. \textbf{80}, 885 (2008)}.
\bibitem{Wu2016} Z. Wu, L. Zhang, W. Sun, X.-T. Xu, B.-Z. Wang, S.-C. Ji, Y. Deng, S. Chen, X.-J. Liu, and J.-W. Pan, \textit{Realization of two-dimensional spin-orbit coupling for Bose-Einstein condensates}, \href{https://www.science.org/doi/10.1126/science.aaf6689}{Science \textbf{354}, 83 (2016)}.
\bibitem{Greiner2002} M. Greiner, O. Mandel, T. Esslinger, T. W. H{\"a}nsch, and I. Bloch, \textit{Quantum phase transition from a superfluid to a Mott insulator in a gas of ultracold atoms}, \href{https://doi.org/10.1038/415039a}{Nature \textbf{415}, 39 (2002)}.
\bibitem{Spielman2008} I. B. Spielman, W. D. Phillips, and J. V. Porto, \textit{Condensate Fraction in a 2D Bose Gas Measured across the Mott-Insulator Transition}, \href{https://doi.org/10.1103/PhysRevLett.100.120402}{Phys. Rev. Lett. \textbf{100}, 120402 (2008)}.
\bibitem{Thomas2017} C. K. Thomas, T. H. Barter, T.-H. Leung, M. Okano, G.-B. Jo, J. Guzman, I. Kimchi, A. Vishwanath, and D. M. Stamper-Kurn, \textit{Mean-Field Scaling of the Superfluid to Mott Insulator Transition in a 2D Optical Superlattice}, \href{https://doi.org/10.1103/PhysRevLett.119.100402}{Phys. Rev. Lett. \textbf{119}, 100402 (2017)}.
\bibitem{Jrdens08} R. J{\"o}rdens, N. Strohmaier, K. G\"unter, H. Moritz, and T. Esslinger, \textit{A  Mott insulator of fermionic atoms in an optical lattice}, \href{https://doi.org/10.1038/nature07244}{Nature \textbf{455}, 204 (2008)}.
\bibitem{Schneider08}  U. Schneider, L. Hackerm\"uller, S. Will, Th. Best, I. Bloch, T. A. Costi, R. W. Helmes, D. Rasch, and A. Rosch, \textit{Metallic and insulating phases of repulsively interacting fermions in a 3D optical lattice}, \href{https://www.science.org/doi/10.1126/science.1165449}{Science \textbf{322}, 1520 (2008)}.
\bibitem{Greif13} D. Greif, T. Uehlinger, G. Jotzu, L. Tarruell, and  T. Esslinger, \textit{Short-range quantum magnetism of ultracold fermions in an optical lattice}, \href{https://www.science.org/doi/10.1126/science.1236362}{Science \textbf{340}, 1307 (2013)}.

\bibitem{Hart15} R. A. Hart, P. M. Duarte, T.-L. Yang, X. Liu, T. Paiva, E. Khatami, R. T. Scalettar, N. Trivedi,
D. A. Huse, and R. G. Hulet, \textit{Observation of antiferromagnetic correlations in the Hubbard model with ultracold atoms}, \href{https://doi.org/10.1038/nature14223}{Nature \textbf{519}, 211 (2015)}.

\bibitem{Parsons16} M. F. Parsons, A. Mazurenko, C. S. Chiu, G. Ji, D. Greif and M. Greiner, \textit{Site-resolved measurement of the spin-correlation function in the Fermi-Hubbard model}, \href{https://www.science.org/doi/10.1126/science.aag1430}{Science \textbf{353}, 1253 (2016)}.
\bibitem{Cheuk16} L. W. Cheuk, M. A. Nichols, K. R. Lawrence, M. Okan, H. Zhang, E. Khatami, N. Trivedi, T. Paiva, M. Rigol, and M. W. Zwierlein, \textit{Observation of spatial charge and spin correlations in the 2D Fermi-Hubbard model}, \href{https://www.science.org/doi/10.1126/science.aag3349}{Science \textbf{353}, 1260 (2016)}.
\bibitem{Koepsell2021} J. Koepsell, D. Bourgund, P. Sompet, S. Hirthe, A. Bohrdt, Y. Wang, F. Grusdt, E. Demler, G. Salomon,
C. Gross, and I. Bloch, \textit{Microscopic evolution of doped Mott insulators from polaronic metal to Fermi liquid}, \href{https://www.science.org/doi/10.1126/science.abe7165}{Science \textbf{374}, 82 (2021)}.
\bibitem{Boll16} M. Boll, T. A. Hilker, G. Salomon, A. Omran, J. Nespolo, L. Pollet, I. Bloch, and C. Gross, \textit{Spin-and density-resolved microscopy of antiferromagnetic correlations in Fermi-Hubbard chains}, \href{https://www.science.org/doi/10.1126/science.aag1635}{Science \textbf{353}, 1257 (2016)}.
\bibitem{Brown17} P. T. Brown, D. Mitra, E. Guardado-Sanchez, P. Schau\ss{}, S. S. Kondov, E. Khatami, T. Paiva, N. Trivedi, D. A. Huse, and W. S. Bakr, \textit{Spin-imbalance in a 2D Fermi-Hubbard system}, \href{https://www.science.org/doi/10.1126/science.aam7838}{Science \textbf{357}, 1385 (2017)}.
\bibitem{Arovas2022} D. P. Arovas, E. Berg, S. A. Kivelson, S. Raghu, \textit{The Hubbard Model}, \href{https://doi.org/10.1146/annurev-conmatphys-031620-102024}{Annual Review of Condensed Matter Physics \textbf{13}, 239 (2022)}.
\bibitem{Scalettar89} R. T. Scalettar, E. Y. Loh, J. E. Gubernatis, A. Moreo, S. R. White, D. J. Scalapino, R. L. Sugar, and E. Dagotto. \textit{Phase diagram of the two-dimensional negative-$U$ Hubbard model}, \href{https://doi.org/10.1103/PhysRevLett.62.1407}{Phys. Rev. Lett. \textbf{62}, 1407 (1989)}.
\bibitem{Kyung01} B. Kyung, S. Allen, and A.-M. S. Tremblay, \textit{Pairing fluctuations and pseudogaps in the attractive Hubbard model}, \href{https://doi.org/10.1103/PhysRevB.64.075116}{Phys. Rev. B \textbf{64}, 075116 (2001)}.
\bibitem{Honerkamp2004} C. Honerkamp, and W. Hofstetter, \textit{Ultracold Fermions and the SU(N) Hubbard Model}, \href{https://doi.org/10.1103/PhysRevLett.92.170403}{Phys. Rev. Lett.  \textbf{92}, 170403 (2004)}.
\bibitem{Mondaini2015}  R. Mondaini, P. Nikoli\'{c}, and M. Rigol, \textit{Mott-insulator-to-superconductor transition in a two-dimensional superlattice}, \href{https://doi.org/10.1103/PhysRevA.92.013601}{Phys. Rev. A \textbf{92}, 013601 (2015)}.
\bibitem{Cocchi16} E. Cocchi, L. A. Miller, J. H. Drewes, M. Koschorreck, D. Pertot, F. Brennecke, and M. K{\"o}hl, \textit{Equation of state of the two-dimensional Hubbard model}, \href{https://doi.org/10.1103/PhysRevLett.116.175301}{Phys. Rev. Lett. \textbf{116}, 175301 (2016)}.
\bibitem{Strohmaier07}  N. Strohmaier, Y. Takasu, K. G{\"u}nter, R. J{\"o}rdens, M. K{\"o}hl, H. Moritz, and T. Esslinger, \textit{Interaction-controlled transport of an ultracold Fermi gas}, \href{https://doi.org/10.1103/PhysRevLett.99.220601}{Phys. Rev. Lett. \textbf{99}, 220601 (2007)}.
\bibitem{Ho04} A. F. Ho, M. A. Cazalilla, and T. Giamarchi, \textit{Quantum simulation of the Hubbard model: the attractive route}, Phys. Rev. A \textbf{79}, 033620 (2009).
\bibitem{Moreo07} A. Moreo, D. J. Scalapino, \textit{Cold attractive spin polarized Fermi lattice gases and the doped positive $U$ Hubbard model}, \href{https://doi.org/10.1103/PhysRevLett.98.216402}{Phys. Rev. Lett. \textbf{98}, 216402 (2007)}.
\bibitem{Gukelberger16} J. Gukelberger, S. Lienert, E. Kozik, L. Pollet, and M. Troyer, \textit{Fulde-Ferrell-Larkin-Ovchinnikov pairing as leading instability on the square lattice}, \href{https://doi.org/10.1103/PhysRevB.94.075157}{Phys. Rev. B \textbf{94}, 075157 (2016)}.
\bibitem{Paiva04} T. Paiva, R. R. dos Santos, R. T. Scalettar, and P. J. H. Denteneer, \textit{Critical temperature for the two-dimensional attractive Hubbard model}, \href{https://doi.org/10.1103/PhysRevB.69.184501}{Phys. Rev. B \textbf{69}, 184501 (2004)}.
\bibitem{Shenoy2008}  V. B. Shenoy, \textit{Phase diagram of the attractive Hubbard model with inhomogeneous interactions}, \href{https://doi.org/10.1103/PhysRevB.78.134503}{Phys. Rev. B \textbf{78}, 134503 (2008)}.

\bibitem{Mitra18} D. Mitra, P. T. Brown, E. Guardado-Sanchez, S. S. Kondov, T. Devakul, D. A. Huse, P. Schau\ss{}, and W. S. Bakr, \textit{Quantum gas microscopy of an attractive Fermi-Hubbard system}, \href{https://doi.org/10.1038/nphys4297}{Nat. Phys. \textbf{14}, 173 (2018)}.
\bibitem{Peter20} P. T. Brown, E. Guardado-Sanchez, B. M. Spar, E. W. Huang, T. P. Devereaux, and W. S. Bakr, \textit{Angle-resolved photoemission spectroscopy of a Fermi-Hubbard system}, \href{https://doi.org/10.1038/s41567-019-0696-0}{Nat. Phys. \textbf{16}, 26 (2020)}.
\bibitem{Hackermuller10} L. Hackerm{\"u}ller, U. Schneider, M. Moreno-Cardoner, T. Kitagawa, T. Best, S. Will, E. Demler, E. Altman, I. Bloch, and B. Paredes, \textit{Anomalous Expansion of Attractively Interacting Fermionic Atoms in an Optical Lattice}, \href{https://www.science.org/doi/10.1126/science.1184565}{Science \textbf{327}, 1621 (2010)}.
\bibitem{Gall2020} M. Gall, C. F. Chan, N. Wurz, and M. K{\"o}hl, \textit{Simulating a Mott Insulator Using Attractive Interaction}, \href{https://doi.org/10.1103/PhysRevLett.124.010403}{Phys. Rev. Lett. \textbf{124}, 010403 (2020)}.
\bibitem{Schneider12} U. Schneider, L. Hackerm{\"u}ller, J. P. Ronzheimer, S. Will, S. Braun, T. Best, I. Bloch, E. Demler, S. Mandt, D. Rasch, and A. Rosch, \textit{Fermionic transport and out-of-equilibrium dynamics in a homogeneous Hubbard model with ultracold atoms}, \href{https://doi.org/10.1038/nphys2205}{Nat. Phys. \textbf{8}, 213 (2012)}.
\bibitem{Hartke2023} T. Hartke, B. Oreg, C. Turnbaugh, N. Jia, M. Zwierlein, \textit{Direct observation of nonlocal fermion pairing in an attractive Fermi-Hubbard gas}, \href{https://www.science.org/doi/10.1126/science.ade4245}{Science \textbf{381}, 82 (2023)}.

\bibitem{Vitali2020} E. Vitali, P. Kelly, A. Lopez, G. Bertaina, and D. E. Galli, \textit{Dynamical structure factor of a fermionic supersolid on an optical lattice}, \href{https://doi.org/10.1103/PhysRevA.102.053324}{Phys. Rev. A \textbf{102}, 053324 (2020)}.
\bibitem{Zhao2023} H. Zhao, R. Han, L. Qin, F. Yuan, and P. Zou, \textit{Universal pairing gap measurement proposal by dynamical excitations in 2D doped attractive Fermi-Hubbard model with spin-orbit coupling}, \href{https://doi.org/10.1103/PhysRevA.110.063326}{Phys. Rev. A \textbf{110}, 063326 (2024)}.

\bibitem{Liu2004} X.-J. Liu, H. Hu, A. Minguzzi, and M. P. Tosi,\textit{
Collective oscillations of a confined Bose gas at finite temperature
in the random-phase approximation}, \href{https://doi.org/10.1103/PhysRevA.69.043605}{Phys. Rev. A \textbf{69}, 043605 (2004)}.
\bibitem{He2016} L. He, \textit{Dynamic density and spin responses of a superfluid Fermi gas in the BCS-BEC crossover: Path integral formulation and pair fluctuation theory}, \href{https://www.sciencedirect.com/science/article/pii/S0003491616301312}{Ann. Phys. \textbf{373}, 470 (2016)}.
\bibitem{Ganesh2009} R. Ganesh, A. Paramekanti, and A. A. Burkov, \textit{Collective modes and superflow instabilities of strongly correlated Fermi superfluids},
\href{https://doi.org/10.1103/PhysRevA.80.043612}{Phys. Rev. A \textbf{80}, 043612 (2009)}.
\bibitem{Han2022} R. Han, F. Yuan, and H. Zhao, \textit{Single-particle excitations and metal-insulator transition of ultracold Fermi atoms in one-dimensional optical lattice with spin-orbit coupling}, \href{https://iopscience.iop.org/article/10.1209/0295-5075/ac39ec}{Europhys. Lett. \textbf{139}, 25001 (2022)}.
\bibitem{Nocera2016} A. Nocera, N. D. Patel, J. Fernandez-Baca, E. Dagotto, and G. Alvarez, \textit{Magnetic excitation spectra of strongly correlated quasi-one-dimensional systems: Heisenberg versus Hubbard-like behavior}, \href{https://doi.org/10.1103/PhysRevB.94.205145}{Phys. Rev. B \textbf{94}, 205145 (2016)}.
\bibitem{Zhang1990} Shoucheng Zhang, \textit{Pseudospin Symmetry and New Collective Modes of the Hubbard Model}, \href{https://doi.org/10.1103/PhysRevLett.65.120}{Phys. Rev. Lett. {\bf 65}, 120 (1990)}.
\bibitem{Qin2022} M. Qin, T. Sch{\"a}fer, S. Andergassen, P. Corboz, and E. Gull, \textit{The Hubbard Model: A Computational Perspective},
\href{https://doi.org/10.1146/annurev-conmatphys-090921-033948}{Annual Review of Condensed Matter Physics \textbf{13}, 275 (2022)}.
\bibitem{Belkhir1994} L. Belkhir and M. Randeria, \textit{Crossover from Cooper pairs to composite bosons: A generalized RPA analysis of collective excitations}, \href{https://doi.org/10.1103/PhysRevB.49.6829}{Phys. Rev. B \textbf{49}, 6829 (1994)}.

\bibitem{Schafer2021} T. Sch\"afer, N. Wentzell, F. \ifmmode \check{S}\else \v{S}\fi{}imkovic, Y.-Y. He, C. Hille, M. Klett, C. J. Eckhardt, B. Arzhang,  V. Harkov, F. M. Le R\'egent, A. Kirsch, Y. Wang, A. J. Kim, E. Kozik, E. A. Stepanov, A. Kauch, S. Andergassen, P. Hansmann, D. Rohe, Y. M. Vilk, J. P. F. LeBlanc, S. Zhang, A.-M. S. Tremblay, M. Ferrero, O. Parcollet, and A. Georges, \textit{Tracking the Footprints of Spin Fluctuations: A MultiMethod, MultiMessenger Study of the Two-Dimensional Hubbard Model}, \href{https://doi.org/10.1103/PhysRevX.11.011058}{Phys. Rev. X \textbf{11}, 011058 (2021)}.
\bibitem{Hafermann2014} H. Hafermann, E. G. C. P. van Loon, M. I. Katsnelson, A. I. Lichtenstein, and O. Parcollet, \textit{Collective charge excitations of strongly correlated electrons, vertex corrections, and gauge invariance}, \href{https://doi.org/10.1103/PhysRevB.90.235105}{Phys. Rev. B \textbf{90}, 235105 (2014)}.
\bibitem{Feng2015} S. Feng, L. Kuang, and H. Zhao,  \textit{Electronic structure of cuprate superconductors in a full charge-spin recombination scheme}, \href{https://doi.org/10.1016/j.physc.2015.06.017}{Physica C \textbf{517}, 5 (2015)}.
\bibitem{Anderson2004} P. W. Anderson, P. A. Lee, M. Randeria, T. M. Rice, N. Trivedi and F. C. Zhang, \textit{The physics behind high-temperature superconducting cuprates: the 'plain vanilla' version of RVB}, \href{https://iopscience.iop.org/article/10.1088/0953-8984/16/24/R02}{J. Phys.: Condens. Matter \textbf{16}, R755 (2004)}.
 \bibitem{Crameri2018} F. Crameri, \textit{Geodynamic diagnostics, scientific visualisation and stagLab 3.0}, \href{https://doi.org/10.5194/gmd-11-2541-2018}{Geosci. Model Dev. \textbf{11}, 2541, (2018)}.
\end{thebibliography}
\end{document}